\documentclass[]{emulateapj}

\usepackage{natbib}
\usepackage{bm}
\usepackage{amsmath}
\shorttitle{}
\shortauthors{}

\begin{document}

\title{The Spatially Uniform Spectrum of the Fermi Bubbles: the Leptonic AGN Jet Scenario}

\author{H.-Y.\ K.\ Yang\altaffilmark{1,2} and
M.\ Ruszkowski \altaffilmark{3}} 
\altaffiltext{1}{Einstein Fellow}
\altaffiltext{2}{Department of Astronomy and Joint Space-Science Institute, University of Maryland, College Park, MD}
\altaffiltext{3}{Department of Astronomy, University of Michigan, Ann Arbor, MI}
\email{Email: hsyang@astro.umd.edu}

\begin{abstract}

The Fermi bubbles are among the most important findings of the Fermi Gamma-ray Space Telescope; however, their origin is still elusive. One of the unique features of the bubbles is that their gamma-ray spectrum, including a high-energy cutoff at $\sim110$ GeV and the overall shape of the spectrum, is nearly spatially uniform. The high-energy spectral cutoff is suggestive of a leptonic origin due to synchrotron and inverse-Compton cooling of cosmic-ray (CR) electrons; however, even for a leptonic model, it is not obvious why the spectrum should be spatially uniform. In this work, we investigate the bubble formation in the leptonic jet scenario using a new CRSPEC module in FLASH that allows us to track the evolution of CR spectrum during the simulations. We show that the high-energy cutoff is caused by fast electron cooling near the Galactic center (GC) when the jets were launched. Afterwards, the dynamical timescale becomes the shortest among all relevant timescales, and therefore the spectrum is essentially advected with only mild cooling losses. This could explain why the bubble spectrum is nearly spatially uniform: the CRs from different parts of the bubbles as seen today all share the same origin near the GC at early stage of the bubble expansion. We find that the predicted CR spatial and spectral distribution can simultaneously match the normalization, spectral shape, and high-energy cutoff of the observed gamma-ray spectrum and their spatial uniformity, suggesting that past AGN jet activity is a likely mechanism for the formation of the Fermi bubbles.

\end{abstract}



\section{Introduction}

The {\it Fermi} bubbles, two giant bubbles extending 50 degrees above and below the Galactic center (GC), are among the most important findings of the {\it Fermi} Gamma-ray Space Telescope \citep{Su10, Ackermann14, Narayanan17}. The observed gamma-ray bubbles have many unique charateristics, including the spatially uniform hard spectrum, nearly flat surface brightness distribution, sharp edges, and smooth surface. The bubbles are also spatially coincident with features in other wavelengths, such as the microwave haze \citep{Finkbeiner04, PlanckHaze}, X-ray properties of the Galactic halo \citep[e.g.,][]{Snowden97, BlandHawthorn03, Kataoka13, Tahara15, Kataoka15, Miller16}, UV absorption lines \citep{Fox15, Bordoloi17}, and polarized lobes \citep{Carretti13}. Because of the proximity, the spatially resolved, multi-wavelength observational data provides unprecedented opportunities for studying the physical origin of the bubbles as well as cosmic ray (CR) propagation, Galactic magnetic field, and past activity at the GC. 

Many theoretical models have been proposed to explain the formation of the bubbles. The hard spectrum of the observed bubbles implies that the CRs, if transported from the GC, must reach large distances before they have time to cool. This consideration gives a constraint on the age of the bubbles to be a few Myr if the gamma rays are produced by CR electrons (CRe, i.e., the leptonic model). In order to satisfy the age constraint, the theories can be divided into three categories: hadronic transport \citep[e.g.,][]{Crocker11, Mou14, Crocker15}, leptonic transport \citep[e.g.,][the latter two are abbreviated as Y12 and Y13, respectively]{Guo12a, Guo12b, Yang12, Yang13}, and in-situ acceleration models \citep[e.g.,][]{Mertsch11, Cheng11, Cheng15, Sarkar15, Sasaki15}. In the hadronic transport models, the gamma rays are generated by inelastic collisions between CR protons (CRp) and thermal nuclei. The CRp are produced at the GC by nuclear starburst or activity of the central active galactic nucleus (AGN), and they are subsequently transported via starburst or AGN driven winds. The hadronic models can successfully reproduce the properties of the observed gamma-ray bubbles; however, to model the microwave haze is nontrivial \citep{Ackermann14} and requires an additional population of primary CRe \citep{Crocker15}. In the leptonic transport models, CRe are injected at the GC via past jet activity of the central supermassive black hole (SMBH) and transported by fast AGN jets. Previous simulations have shown that the bubbles can be inflated within a few Myr \citep[][Y12]{Guo12a}. Also, the key features of the gamma-ray bubbles as well as the microwave and polarization signatures are in good agreements with the observational data (Y12, Y13). Some observational studies of the thermal and kinematic properties of the Galactic halo \citep[e.g.,][]{Kataoka13, Sarkar17} suggest that the {\it Fermi} bubbles are triggered by milder outflows, which could potentially be in tension with the jet model. However, more data is needed to draw a conclusion because the Galactic halo in the vicinity of the bubbles is extremely complex \citep{Tahara15, Kataoka15}. Also, there are discrepancies among observationally derived kinematics \citep[e.g.,][]{Sarkar17, Bordoloi17}, possibly due to modeling uncertainties such as the assumptions of geometry and injection pattern of the outflows. For the in-situ acceleration models, CRs are assumed to be produced by shocks or turbulence near the edges of the bubbles. Although these models could bypass the age constraints, it has been challenging for the simplest models to produce the flat gamma-ray intensity profile \citep{Mertsch11, Cheng11} as well as the microwave haze emission \citep{Fujita14, Cheng15}.     


One unique and important feature of the observed bubbles that has not been investigated in detail is the spatially uniform gamma-ray spectrum. \cite{Ackermann14} showed that the bubble spectrum can be well fit by a power law with an exponential cutoff at $\sim 110$\ GeV. Remarkably, both the shape of the spectrum and the cutoff energy are almost independent of Galactic latitude \citep[see also][]{Narayanan17}. The high-energy cutoff is suggestive of a leptonic origin because CRe can cool more easily due to synchrotron and inverse-Compton (IC) energy losses. However, even for a leptonic model, it is unclear why the spectrum should be spatially uniform.   


In Y12 and Y13, we investigated the leptonic AGN jet scenario using three-dimensional (3D) magnetohydrodynamic (MHD) simulations including relevant cosmic-ray (CR) physics. As mentioned above, the leptonic jet model is a promising mechanism for explaining the origin of the bubbles as it is the simplest model that could simultaneously explain the gamma-ray bubbles and the microwave haze. However, in the previous works, the CRs are treated as a single fluid without distinguishing their energies, and therefore comparisons with observations have to rely on assumptions of the CR spectrum. In this study, we implement a new CRSPEC module in the FLASH code \citep{Flash, Dubey08} that could handle CRs of different energy channels and follow their spectral evolution {\it on-the-fly} during the simulations. We apply it to simulate the spectral evolution of the CRs within {\it Fermi} bubbles and generate the gamma-ray spectrum {\it self-consistently}. Our objectives are to answer the questions: (1) what physical mechanisms are responsible for the $\sim 110$\ GeV cutoff in the observed gamma-ray spectrum? (2) why is the bubble spectrum spatially uniform, including both the overall spectral shape and the cutoff energy?  

The structure of the paper is as follows. We first discuss expectations of the CR spectra as hinted by the gamma-ray data in \S~\ref{sec:hints}. In \S~\ref{sec:method} we outline the simulation setup and describe key aspects of the CRSPEC code. In \S~\ref{sec:result} we present results from the simulations, including the simulated distribution of CR energies (\S~\ref{sec:ecr}), general spectral evolution of the bubbles (\S~\ref{sec:evolution}), the spatial dependence of the spectrum (\S~\ref{sec:spatial}), and constraints on the AGN jet speed, magnetic field strength, and energy density of the interstellar radiation field (ISRF) derived from our model (\S~\ref{sec:constraint}). Finally we summarize our findings in \S~\ref{sec:conclusion}.    


\section{Hints from the observed spectrum}
\label{sec:hints}

The observed gamma-ray spectrum of the {\it Fermi} bubbles is strikingly latitude independent, including its shape and the high-energy cutoff (e.g., Fig.\ 33 in \cite{Ackermann14}). The observed bubble spectrum can be best fit by a power law with an exponential cutoff term $\exp(-E/E_{\rm cut}$), where $E_{\rm cut} \sim 110$ GeV. This energy scale represents where the gamma-ray spectrum has a turnover; however, there could still be gamma rays generated beyond this energy (see Figure \ref{fig:spectra}). In order to connect with the energies of the underlying CRs more directly, hereafter we define a ``maximum energy of the observed gamma rays" as $E_{\rm max,obs}$, which relates to $E_{\rm cut}$ by $\exp(-E_{\rm max,obs}/E_{\rm cut})=0.1$ (i.e., the energy scale where the gamma-ray intensity is dimmer by a factor of 10). Given $E_{\rm cut} = 110$\ GeV, we have $E_{\rm max,obs} \sim 250$\ GeV. In this section we show that the data alone can readily provide some clues about the underlying CR spectra and their latitude dependence. Specifically, for a given latitude bin, the shape of the gamma-ray spectra can inform the characteristic energy of the CRe, and the observed cutoff energy is related to the maximum energy of the CRe.   

In the leptonic scenario, the gamma rays originate from IC scattering of the ISRF by CRe. The spectrum of the up-scattered photons per one electron of Lorentz factor $\gamma$
is given by \cite{Blumenthal70},

\begin{eqnarray}
\frac{dN}{dE_\gamma dE_{\rm ph} dt} &=& \frac{3}{4} \sigma_T c \frac{(m_e c^2)^2}{E_{\rm ph} E_e^2} (2q \log q + (1+2q)(1-q) \nonumber \\ 
&+& 0.5(1-q)(\Gamma q)^2 / (1+\Gamma q)) n(E_{\rm ph}),\\
\Gamma &=& 4E_{\rm ph}E_{\rm e}/(m_e c^2)^2 = 4 \gamma E_{\rm ph}/(m_e c^2), \\
q &=& \frac{E_\gamma}{E_{\rm e}} \frac{1}{\Gamma(1-E_\gamma/E_{\rm e})},\\
E_{\rm ph} &<& E_\gamma < E_{\rm e} \Gamma/(1+\Gamma),
\end{eqnarray}
where $E_{\rm ph}$ is the initial photon energy, $\gamma = E_{\rm e}/(m_e c^2)$ is the Lorentz factor of the CR electron, $E_\gamma$ is the energy of the up-scattered gamma-ray photon, and $n(E_{\rm ph})$ is the energy distribution of the photon number density.  

In the Thomson limit ($\Gamma \ll 1$), the average energy of the up-scattered photons is given by
\begin{equation}
\langle E_\gamma \rangle = (4/3) \gamma^2 \langle E_{\rm ph} \rangle. 
\label{eq:egamma}
\end{equation}
In the Klein-Nishina (KN) limit ($\Gamma \gg 1$), almost all the energy of the CR electron is carried away by the up-scattered photons, i.e., $\langle E_\gamma \rangle \sim \langle E_{\rm e} \rangle$. Assuming the CR spectrum is a power law with spectral index $\alpha$, it can be shown that the spectral index of the up-scattered gamma-ray photons is $(\alpha+1)/2$ in the Thomson limit, and approximately $\alpha+1$ in the KN limit \citep{Blumenthal70}. The observed bubble spectrum is best fit by a power-law CR distribution with spectral index of $\sim 2$, and therefore one may expect to see changes in the spectral indices from 1.5 to 3 in the gamma-ray spectrum as the IC scattering goes from the Thomson limit to the KN regime. 

The observed spectrum of the bubbles is nearly latitude independent, characterized by a broad bump that roughly peaks around $E_{\rm bump} \sim 10$\ GeV (see Figure \ref{fig:spectra}). This is not straightforward to obtain because the ISRF is dominated by the cosmic microwave background (CMB) at high latitudes and optical starlight at low latitudes. If the underlying CRe had identical spectra across all latitudes, then the resulting gamma-ray spectra would peak at lower energies at higher latitudes. This implies that the {\it average} energy of the CR population must be latitude {\it dependent}. In fact one could estimate the average energy of CRe for different latitude bins in the Thomson limit using Eq.\ \ref{eq:egamma} because $\Gamma < 1$ for $E_{\gamma} \sim 10$\ GeV and $E_{\rm ph} < 10$\ eV. 

For high latitudes (e.g., $b=40^\circ - 60^\circ$), the ISRF peaks at $\langle E_{\rm ph} \rangle \sim 7\times 10^{-4}$\ eV for CMB photons. It would therefore require an average CR energy of $\langle E_{\rm e} \rangle \sim 2$\ TeV in order to produce a $\sim 10$\ GeV bump. For intermediate latitudes ($b=20^\circ - 40^\circ$), the intensity of the ISRF is more uniform across all wavelengths. Assuming the gamma-ray bump primarily comes from infrared (IR) photons ($E_{\rm ph} \sim 10^{-2}$\ eV), one would obtain $\langle E_{\rm e} \rangle \sim 200$\ GeV. Similarly, the average CR energy can be estimated to be $\langle E_{\rm e} \rangle \sim 20$\ GeV at low latitudes where optical light (assuming $\langle E_{\rm ph} \rangle \sim 5$\ eV) dominates the ISRF. Therefore, generally speaking, the spatially uniform spectra of the {\it Fermi} bubbles require that the average energy of CRe to be higher at higher latitudes. The exact magnitude of the energy gradient, though, may be different from the above estimate because the observed gamma-ray bump is broad and $E_{\rm bump}$ does not have to be close to 10\ GeV. In fact, the gradient of CR energies should be smaller in order to be consistent with the maximum energy of CRe as estimated below.    

On the other hand, the maximum energy of the observed gamma-ray spectrum, $E_{\rm max,obs}$, comes from up-scattered optical light in the ISRF and provides information about the maximum energy of the underlying CR electron population, $E_{\rm max}$. For optical photons ($\langle E_{\rm ph} \rangle \sim 5$\ eV), the IC scattering would be in the KN limit for CRe with energies greater than $\sim 100$\ GeV. Therefore, for high and intermediate latitudes, $E_{\rm max} \gtrsim E_{\rm max,obs} \sim 250$\ GeV as it is in the KN regime. For low latitudes, the average CR energy is smaller and hence the formula in the Thomson limit applies. Assuming $E_{\rm max,obs} = 250$\ GeV, Eq.\ \ref{eq:egamma} gives $E_{\rm max} \sim 100$\ GeV. The observed cutoff energy is almost independent of latitudes, with a slight tendency of higher $E_{\rm max,obs}$ for higher latitudes (Figure 33 of \cite{Ackermann14}). Therefore, the estimates above imply that $E_{\rm max}$ is also nearly spatially uniform, on the order of a few hundred GeV, and may be somewhat greater at higher latitudes. Our simple estimates of the CR electron cutoff energy for different latitudes are consistent with best-fit values to the observed gamma-ray spectrum obtained by \cite{Narayanan17}.


\section{Methodology}
\label{sec:method}


We simulate the spectral evolution of the {\it Fermi} bubbles in the leptonic AGN jet scenario using 3D hydrodynamic simulations including CRs. The simulation setup is essentially identical to that of Y12 and Y13, to which we refer the readers for details including the initial conditions for the Galactic halo as well as parameters for the AGN jets. Here we briefly summarize our approach and emphasize the differences from the previous works. 

The simulations are performed using the adaptive-mesh-refinement (AMR) code FLASH \citep{Flash, Dubey08}. Same as in Y12, the CRs are injected at the GC during a short (0.3\ Myr), active phase of the central SMBH about 1.2\ Myr ago and then the CRs are advected with the AGN jets. CR diffusion is omitted in the present work since it is slow enough that it only affects the sharpness of the bubble edges but not the overall dynamics and CR distribution (Y12). However, we return to this point in \S~\ref{sec:spatial}. As in Y12, we assume that CRs are scattered by extrinsic turbulence rather than self-excited Alfven waves, and therefore the effects of CR streaming is not included. Since only a small fraction of injected CRs is needed to reproduce the gamma-ray signal (Y13), we assume $3 \times 10^{-3}$ of the injected CRs to be CRe \footnote{We note that the normalization factor is different from that adopted in Y13 because the injected the CR spectrum has a different spectral range.} and follow their spectral evolution due to adiabatic compression, adiabatic expansion, synchrotron losses, and IC losses. The rest of the injected CR energy density \footnote{Though called CR energy density, it is degenerate with the thermal energy density since the dynamics is determined by the total energy density of the jets (Y12).} does not cool and is the dominating component in terms of dynamics. Passively evolving tracer particles are injected along with the jets in order to track the evolution of the bubble spectrum. We note that \cite{Su12} and \cite{Ackermann14} have found substructures in the intensity distribution of the south bubble (i.e., the ``cocoon"), which might be related to a second event of energy injection from the GC. However, in this work we do not consider CR injections from a second AGN outburst in order to avoid introduction of an additional set of jet parameters that are not uniquely constrained. To this end, we refrain ourselves from interpreting the substructures and only focus on the primary bubble emission that has nearly flat intensity distribution.

In Y13, we demonstrated that the magnetic field within the bubbles has to be amplified to values comparable to the ambient field in order to simultaneously produce the microwave haze. Therefore, we do not include magnetic fields in the current simulations but simply assume the default magnetic field distribution as in GALPROP \citep{Strong09}, $|B|=B_0 \exp (-z/z_0) \exp (-R/R_0)$, for the computation of synchrotron losses, where $R$ is the projected radius to the Galaxy's rotational axis. We adopt $z_0=2$\ kpc and $R_0=10$\ kpc, which are best-fit values in the GALPROP model to reproduce the 408 MHz synchrotron radiation in the Galaxy. The normalization of the magnetic field strength $B_0$ is treated as a free parameter and the simulations presented in this paper has a fiducial value of 10 $\mu {\rm G}$. As we will discuss in \S~\ref{sec:result}, the cutoff energy of the gamma-ray spectrum is very sensitive to $B_0$ and therefore could be used to put constraints on the initial conditions. Note that $B_0$ represents the magnetic field strength at the GC right after the initial injection, and therefore does not need to be the same as the field strength as observed today. In fact, $B_0$ is likely smaller than the present observed field strength at the GC \citep[e.g.,][]{Crocker10} because it takes time for the magnetic field within the bubbles to amplify after the initial adiabatic expansion caused by the jets (Y13). 

For IC losses, we adopt the ISRF model from GALPROP v.50 \citep{Strong07} and compute the CR energy losses and gamma-ray emissivity including the KN effects \citep{Jones68}. While there exist other ISRF models that are more general to all spiral galaxies \citep[e.g.,][]{Popescu13}, we chose the GALPROP model because it is calibrated using stellar and dust distributions specific to the Milky Way. Also, it is adopted by all previous studies of the {\it Fermi} bubbles, allowing us to make comparisons to previous results directly. The ISRF model in GALPROP v.50 provides a 3D distribution of photon energy densities for discrete values of $(x,y,z)$ with spacings of 0.1\ kpc. We therefore used 3D linear interpolations to obtain the photon energy densities on our simulation grid. The adopted ISRF decreases away from the GC. Specifically, the values range from $\sim 19\ {\rm eV\ cm}^{-3}$ near the GC to $\sim 1\ {\rm eV\ cm}^{-3}$ at 5 kpc away from the Galactic plane near the rotational axis.

\subsection{Modeling CR spectral evolution}

The core of this work is the newly implemented CRSPEC module in FLASH. Advection of CRs, anisotropic diffusion (though not used in this work), and dynamical coupling between the CRs and the gas are done in the same way as in the previous version (see equations in Y12). But instead of having a single equation for the evolution of the total CR energy density, the CRs are divided into $N_p$ logarithmically spaced momentum bins. Equations are solved for the CR number densities $n_i$ and CR energy densities $e_i$ in each bin with index $i$. The algorithm for solving this set of equations is based on the method for fast cooling electrons in the COSMOCR code \citep{COSMOCR}, and we made modifications in order to handle finite spectral ranges. In the adopted approach, the CR distribution function as a function of momentum, $p$, is approximated with a piece-wise power law, 
\begin{equation}
f(p)=f_i \left( \frac{p}{p_{i-1}} \right)^{-q_i},
\label{eq:fp}
\end{equation}
where $f_i$ and $q_i$ are the normalization and logarithmic slope for the $i$th momentum bin. Fluxes of $n_i$ and $e_i$ across different momentum bins are computed according to adiabatic processes and synchrotron and IC energy losses. For completeness we summarize the details, relevant equations, and test cases in the Appendix.  

We inject CRe with a power-law spectrum from the GC in the beginning of the simulations for a duration of 0.3 Myr. The initial spectrum ranges from 10 GeV to 10 TeV, with a spectral index of $q_i=4.1$. The normalization factor in each bin $f_i$ is chosen so that the energy density of CRe is $3 \times 10^{-3}$ of the total CR energy density of the jets, or equivalently, $7.5\times 10^{-12}\ {\rm erg\ cm}^{-3}$. A representative value for the normalization factor at 10\ GeV is $2.24\times 10^{-4}\ {\rm cm}^{-2}\ {\rm s}^{-1}\ {\rm GeV}^{-1}$. To simulate the spectrum we have five logarithmically spaced momentum bins between 0.1 GeV and 10 TeV. The range is chosen so as to cover the energy shifts of the injected CRs due to the adiabatic and cooling processes. We use a relatively small number of momentum bins in order to minimize computational costs. This is adequate for our application because the CR spectrum, except for the high-energy end, only experiences advection and adiabatic processes, and therefore the spectrum can be well approximated by a power law. In order to accurately simulate the cutoff energy of the CR spectrum at the high-energy end due to fast electron cooling, which is one of the main purposes of the paper, we store extra variables for the minimum and maximum of the CR spectrum ($p_{cutL}$ and $p_{cutR}$, respectively) and track their evolution separately. In order to account for fast synchrotron and IC cooling of CRe accurately, the simulation timestep is set to 0.1 times the cooling timescale. When this timestep is the shortest among all relevant timescales in the simulation, we subcycle over the CR spectral evolution in order to accelerate computations. As a test of our algorithm, we performed a simulation of the {\it Fermi} bubble spectral evolution including only adiabatic processes (synchrotron and IC losses are turned off). We verified that the total CR number density is conserved after the jets are shut off at $t=0.3$ Myr, the spectrum is shifted but the shape is unaltered, and that the total CR energy density distribution at the end of the simulations, $t=1.2$ Myr, is identical to what was obtained using the energy-integrated version of the code as in Y12.        


\section{Results}
\label{sec:result}

\subsection{Distribution of CR energies}
\label{sec:ecr}


\begin{figure*}[tp]
\begin{center}
\includegraphics[scale=0.65]{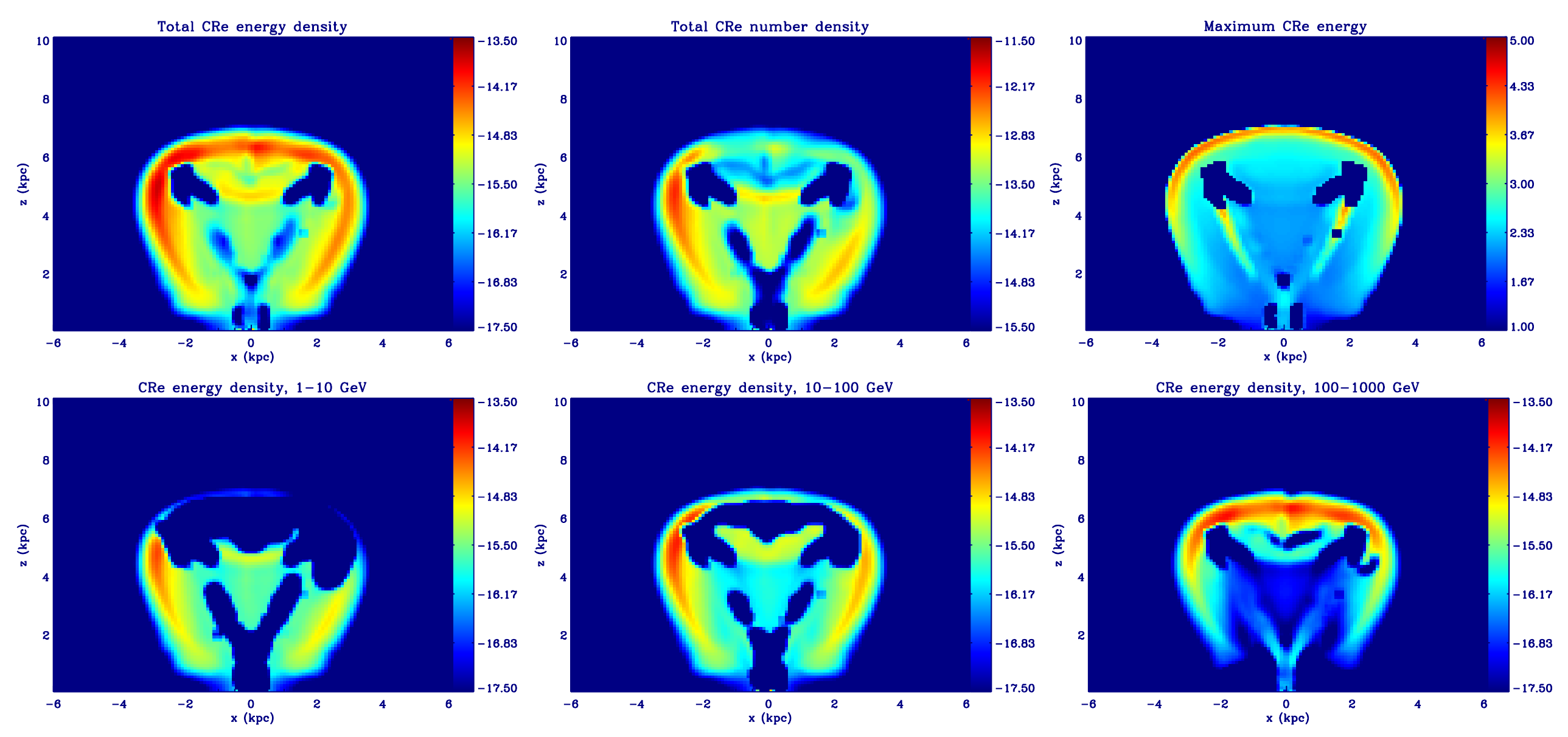} 
\caption{The top row (from left to right) shows slices of the total energy density of simulated CRe (in units of ${\rm erg}\ {\rm cm}^{-3}$), total CR electron number density (in ${\rm cm}^{-3}$), and the maximum CR electron energy (in GeV) at the end of the simulation $t=1.2$\ Myr. The CR electron energy density is further decomposed into different energy channels (bottom row). All quantities are plotted in logarithmic scale.} 
\label{fig:slices}
\end{center}
\end{figure*}

Figure \ref{fig:slices} shows distributions of the CRe at the end of the simulation, $t=1.2$\ Myr, including the total CR electron energy density, total CR electron number density, maximum CR electron energy \footnote{The maximum CR electron energy, $E_{\rm max}$, is solved separately for all cells with nonzero CR energy densities, and therefore the values are computed for an extended region beyond the bubbles where a tiny but nonzero amount of CRs exist due to numerical diffusion. We determined that this is a numerical artifact and therefore only plotted $E_{\rm max}$ for regions where the total energy density of CRe is greater than $10^{-16}\ {\rm erg}\ {\rm cm}^{-3}$, which corresponds to a minuscule level of $\sim 6\times 10^{-5}\ {\rm eV\ cm^{-3}}$.}, and CR electron energy densities in different energy channels. The first and last energy bins (i.e., 0.1-1\ GeV and 1-10\ TeV) are not shown because they contain little amount of CRe at the end of the simulations. 

Because the CRe only contribute to $3\times 10^{-3}$ of the total CR energy density of the jets, they are not dynamically dominant and therefore the overall distribution of the CR electron energy density is similar to that in the adiabatic simulation (Y12, Figure 1). Though some of the details (e.g., structures close to the GC) are slightly different due to cooling of CRe, the main characteristics such as the bubble morphology and the edge-brightened distribution are recovered. We recall that the edge-enhanced CR distribution, which is a result of compression of jet materials during the active phase of the AGN injections, plays a crucial role in reproducing the flat surface brightness distribution of the observed bubbles after line-of-sight projection. In particular, the CR number density near the top of the bubbles must be greater {\it by the right amount} so that, after convolving with the ISRF whose intensity decreases with Galactic latitudes, the projected gamma-ray intensity is almost spatially uniform (Y13). 

One thing to note from Figure \ref{fig:slices} is that the total energy density (upper left panel) and total number density (upper middle panel) of CRe have similar distributions. Since $e_{\rm cr} \sim \langle E_{\rm e} \rangle n_{\rm cr}$, where $\langle E_{\rm e} \rangle$ is the characteristic energy of the CRe, this implies that $\langle E_{\rm e} \rangle$ is largely spatially uniform. This is indeed suggested by the map of maximum CR electron energy $E_{\rm max}$ within the bubbles (upper right panel): while both the CR electron energy and number densities differ by about two orders of magnitudes from minimum to maximum, the variation in $E_{\rm max}$ is relatively small, with $E_{\rm max}$ on the order of a few hundreds of GeV for most regions within the bubbles. 

Although $\langle E_{\rm e} \rangle$ and $E_{\rm max}$ are generally quite spatially uniform, they do exhibit some gradients. The top right panel in Figure \ref{fig:slices} shows that $E_{\rm max}$ varies from $\sim 100$\ GeV at low Galactic latitudes to $\sim 1$\ TeV at high latitudes within the bubbles. There is also a thin shell of very high energy CRs at the edges of the bubbles. However, since they occupy only a very small volume, and that the CR number density is low in this shell, their contribution to the projected CR spectra is negligible (see Figure \ref{fig:crspec}). The gradient in CR energies is also evident by comparing the CR energy densities divided into different energy channels (bottom row in Figure \ref{fig:slices}). The overall uniformness and mild gradient toward higher energies at higher latitudes are consistent with the expectations derived from the observed bubble spectrum (see \S~\ref{sec:hints}). In the following section, we describe the evolution of the CR spectrum in order to understand the final distribution of CR energies as seen in Figure \ref{fig:slices}. 

\subsection{Spectral evolution of the Fermi bubbles}
\label{sec:evolution}

\begin{figure}[tp]
\begin{center}
\includegraphics[scale=0.65]{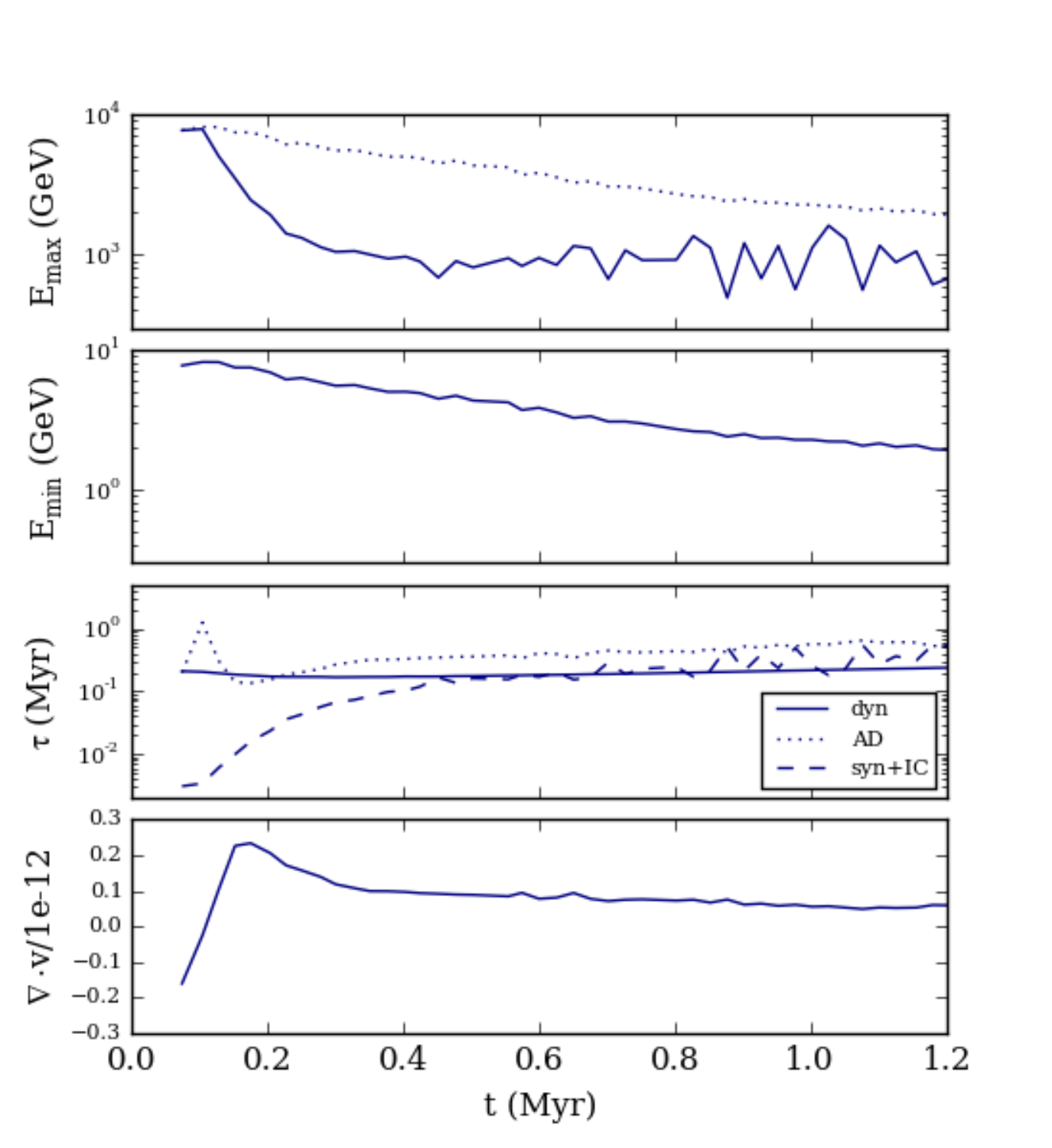} 
\caption{Evolution of one representative tracer particle. Panels from top to bottom show maximum energy of the CR spectrum (the result expected for an adiabatic simulation is overplotted with the dotted line), minimum energy of the CR spectrum, relevant timescales in the simulation (including the dynamical time, timescale for adiabatic compression or expansion, and synchrotron plus IC cooling time), and the divergence of the velocity field.} 
\label{fig:tracer}
\end{center}
\end{figure}

Figure \ref{fig:tracer} shows the evolution of one representative tracer particle that was injected at early stage of the bubble expansion. Panels from top to bottom show the evolution of the maximum energy of the CR spectrum ($E_{\rm max}$), minimum spectral energy ($E_{\rm min}$), relevant timescales, and divergence of the velocity field. The dotted line overplotted in the top panel represents the result from the adiabatic simulation without synchrotron and IC cooling. For this adiabatic case, the change in $E_{\rm max}$ and $E_{\rm min}$ are directly proportional to each other, meaning that the CR spectrum is only shifted without changing the spectral shape. Right after the particle was injected ($t \sim 0.1$\ Myr), there was a brief increase in $E_{\rm min}$ and $E_{\rm max}$ due to adiabatic compression, i.e., $\nabla \cdot {\bm v} < 0$. Afterwards, the CRe propagate outward and the only cooling mechanism is adiabatic expansion ($\nabla \cdot {\bm v} > 0$), and therefore both $E_{\rm min}$ and $E_{\rm max}$ decrease with time monotonically. 

In contrast, the evolution of $E_{\rm max}$ is quite different for the simulation including synchrotron and IC cooling. As shown in the top panel in Figure \ref{fig:tracer} (solid line), the maximum energy of the CRe drops rapidly from the injected energy of 10 TeV to $\sim 1$\ TeV before $t \sim 0.3$\ Myr. This fast change in $E_{\rm max}$ is owing to synchrotron and IC energy losses, as during this early phase of evolution the cooling timescale for synchrotron and IC radiation near the GC is much shorter than all other relevant timescales (see the third panel). After $t\sim 0.4$ Myr, the dynamical timescale ($\tau_{\rm dyn} \equiv (1\ {\rm kpc})/v$) becomes shorter/comparable to the synchrotron and IC cooling timescale, while the timescale for adiabatic processes ($\tau_{\rm AD} \equiv 1/(\nabla \cdot {\bm v}$)) is subdominant. That is, at later stage of the bubble expansion, the CRe experience advection (which does {\it not} cause energy losses) and synchrotron plus IC cooling at the same time (which now occurs on longer timescales compared to the beginning). Therefore, $E_{\rm max}$ only decreases slightly after $t \sim 0.4$\ Myr and reaches a value about 700 GeV at $t=1.2$\ Myr. In other words, the value of $E_{\rm max}$ at the present day is closely related to fast cooling of CRe near the GC at early stage of the bubble expansion when the jets were first launched.

\subsection{Why is the spectrum spatially uniform?}
\label{sec:spatial}


In this section, we provide explanations as to why the gamma-ray spectrum of the {\it Fermi} bubbles is almost spatially uniform, including the maximum energy $E_{\rm max,obs} \sim 250$\ GeV and the overall shape of the spectra for different latitude bins.  

\begin{figure}[tp]
\begin{center}
\includegraphics[scale=0.65]{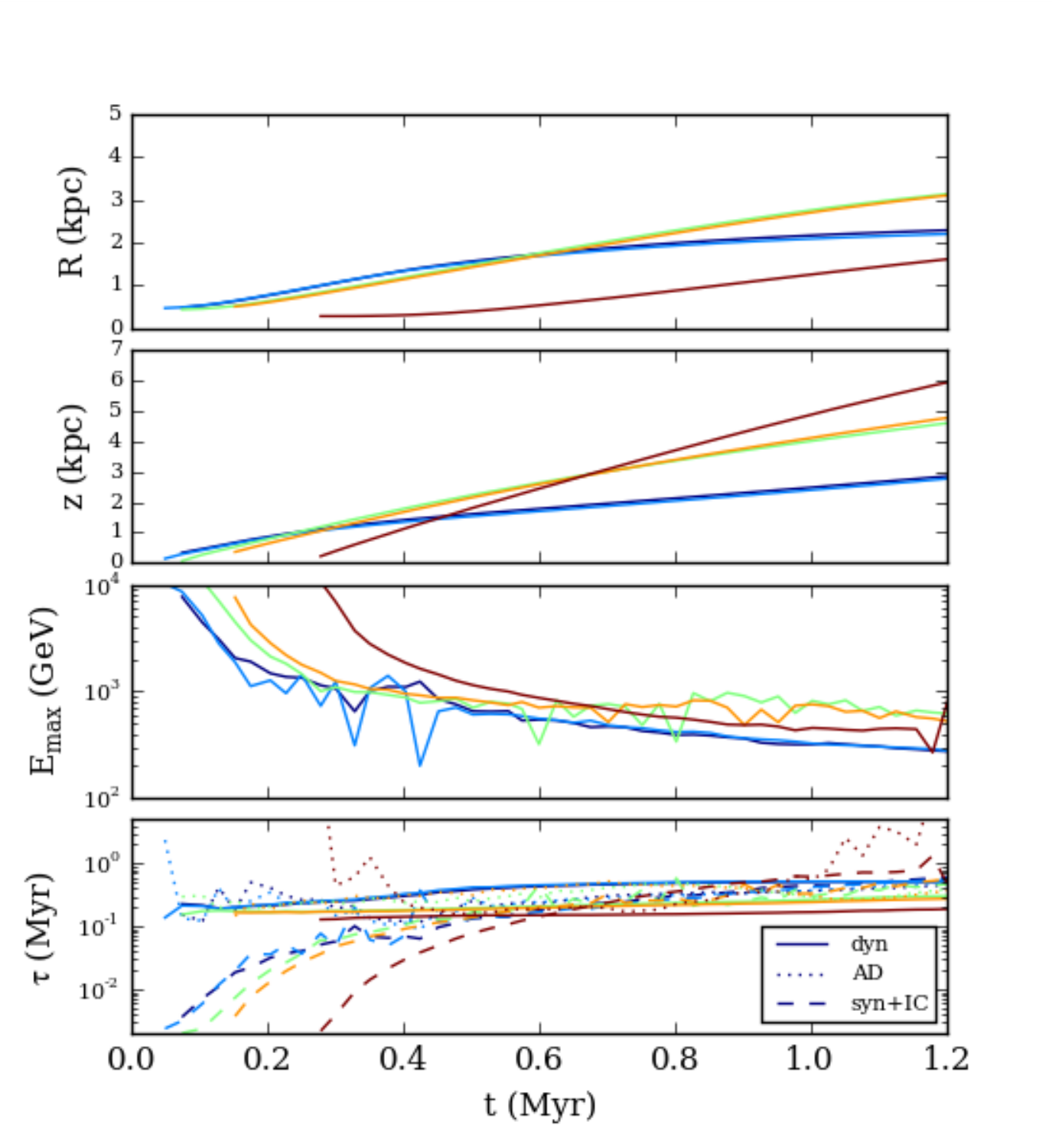} 
\caption{Evolution of a selection of tracers that end up in different locations within the bubbles at $t=1.2$ Myr. Panels from top to bottom show the projected radius to the Galaxy's rotational axis, the vertical distance from the Galactic disk, maximum energy of the CR spectrum, and the relevant timescales in the simulation. } 
\label{fig:tracers}
\end{center}
\end{figure}

Figure \ref{fig:tracers} shows the evolution of a selection of tracers that have different final locations within the bubbles. Their distance to the Galaxy rotational axis, vertical height to the Galactic disk, maximum energy of the CR spectrum, and relevant timescales are plotted in the panels from top to bottom. We find that, although the particles all have distinct trajectories, their evolution of $E_{\rm max}$ is very similar, marked with a fast decay within the early $\sim 0.3$\ Myr after injection and a subsequent mild decrease to several hundreds of GeV at the end of the simulation. Similar to the tracer shown in Figure \ref{fig:tracer}, the CRe encounter significant energy losses due to synchrotron and IC cooling near the GC soon after they are injected with the AGN jets ($\tau_{\rm syn+IC} \ll \tau_{\rm dyn}$ as shown in the bottom panel). Afterwards, the dynamical time of the jets becomes shorter than adiabatic, synchrotron and IC cooling timescales, and hence the CR spectrum is essentially advected with only mild cooling losses. Though somewhat dependent on the time of injection and degree of initial compression, the final value of $E_{\rm max}$ for each tracer particle is similar because the energy scale is mainly set by fast cooling near the GC where the particles all had the same initial conditions. This is the reason why $E_{\rm max}$ is nearly spatially uniform at the present day (top right panel of Figure \ref{fig:slices}).   

\begin{figure*}[tp]
\begin{center}
\includegraphics[scale=0.95]{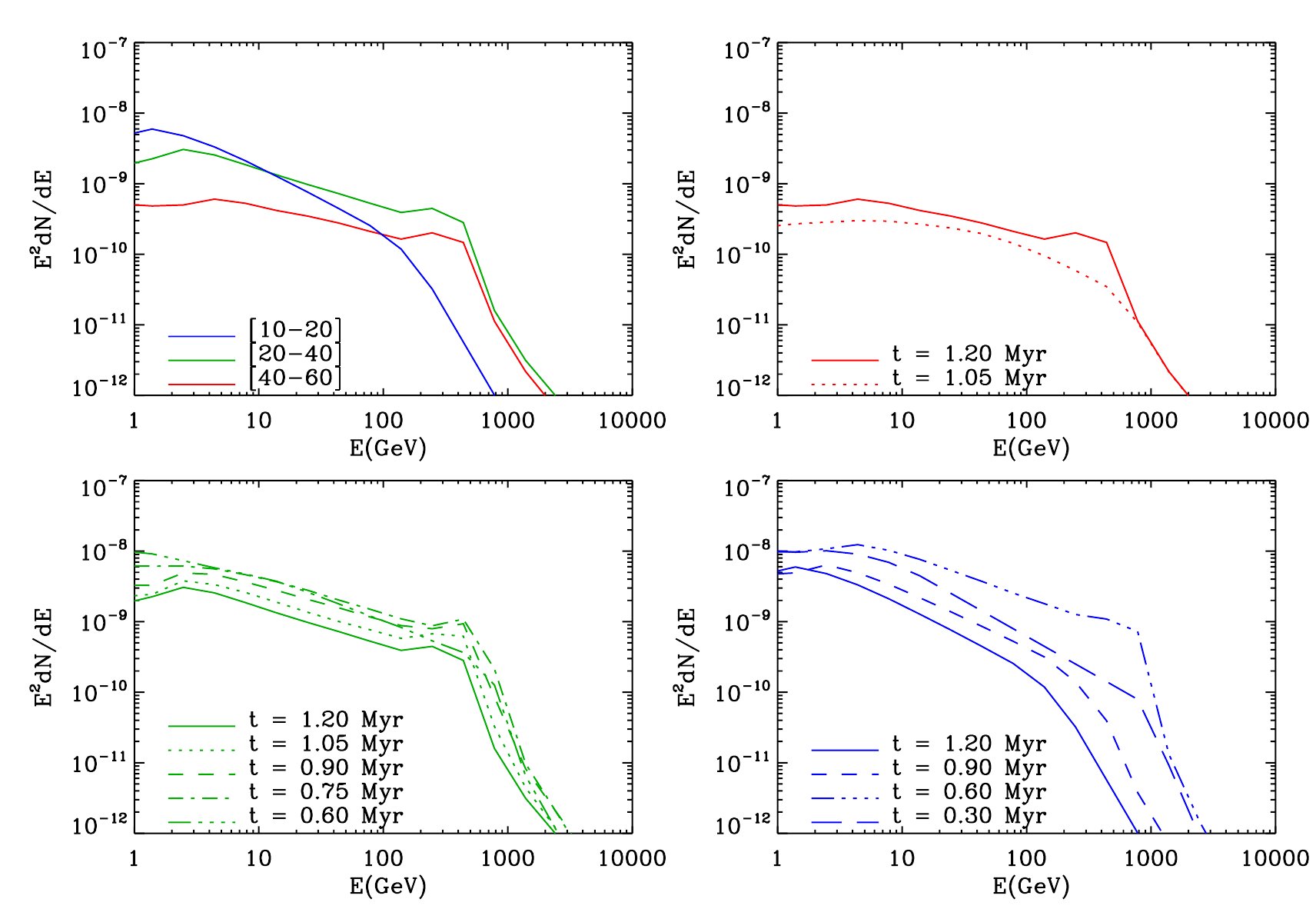} 
\caption{Spectra of the CRe at the present day calculated for a longitude range of $l=[-10^\circ, 10^\circ]$ for different latitude bins (top left panel). The evolution of the spectra for the three latitude bins are shown in the other three panels.} 
\label{fig:crspec}
\end{center}
\end{figure*}

Figure \ref{fig:crspec} shows the spectra of the CRe and their evolution after line-of-sight projections. The red, green, and blue curves represent the CR spectra projected into a longitude range of $l=[-10^\circ, 10^\circ]$ and latitude ranges of $[40^\circ, 60^\circ], [20^\circ, 40^\circ]$, and $[10^\circ, 20^\circ]$, respectively. The top left panel shows the spectra for different latitudes at the present day. This plot confirms our expectation that the maximum energy of the CRe, $E_{\rm max}$, is only mildly varying with latitudes, ranging from $\gtrsim 100$\ GeV at low latitudes to $\sim 1$\ TeV at higher latitudes (see also top right panel of Figure \ref{fig:slices}). Again, the spatial uniformity of $E_{\rm max}$ is resulted from initial fast cooling and subsequent mild adiabatic losses. This process can be seen from the other three panels of Figure \ref{fig:crspec}, in which we plot the CR spectral evolution for the three latitude bins. At early times when the jets just shut off ($t=0.3$\ Myr), only the low latitude bin is populated with CRs (see the long-dashed curve in the lower right panel). Due to the initial cooling, the spectra at this time already showed an exponential cutoff at $\sim 1$\ TeV. Afterwards, the CRs are propagated to higher latitudes, but $E_{\rm max}$ only mildly shifts to lower energies due to adiabatic cooling (the shift is strongest for the lowest latitude bin because the CRe also suffer from stronger synchrotron and IC losses). In terms of the amplitudes of the spectra, in general they decrease with time owing to cooling; this trend is only inverted when the CRs first entered the lowest and highest latitude bins.

\begin{figure*}[tp]
\begin{center}
\includegraphics[scale=0.95]{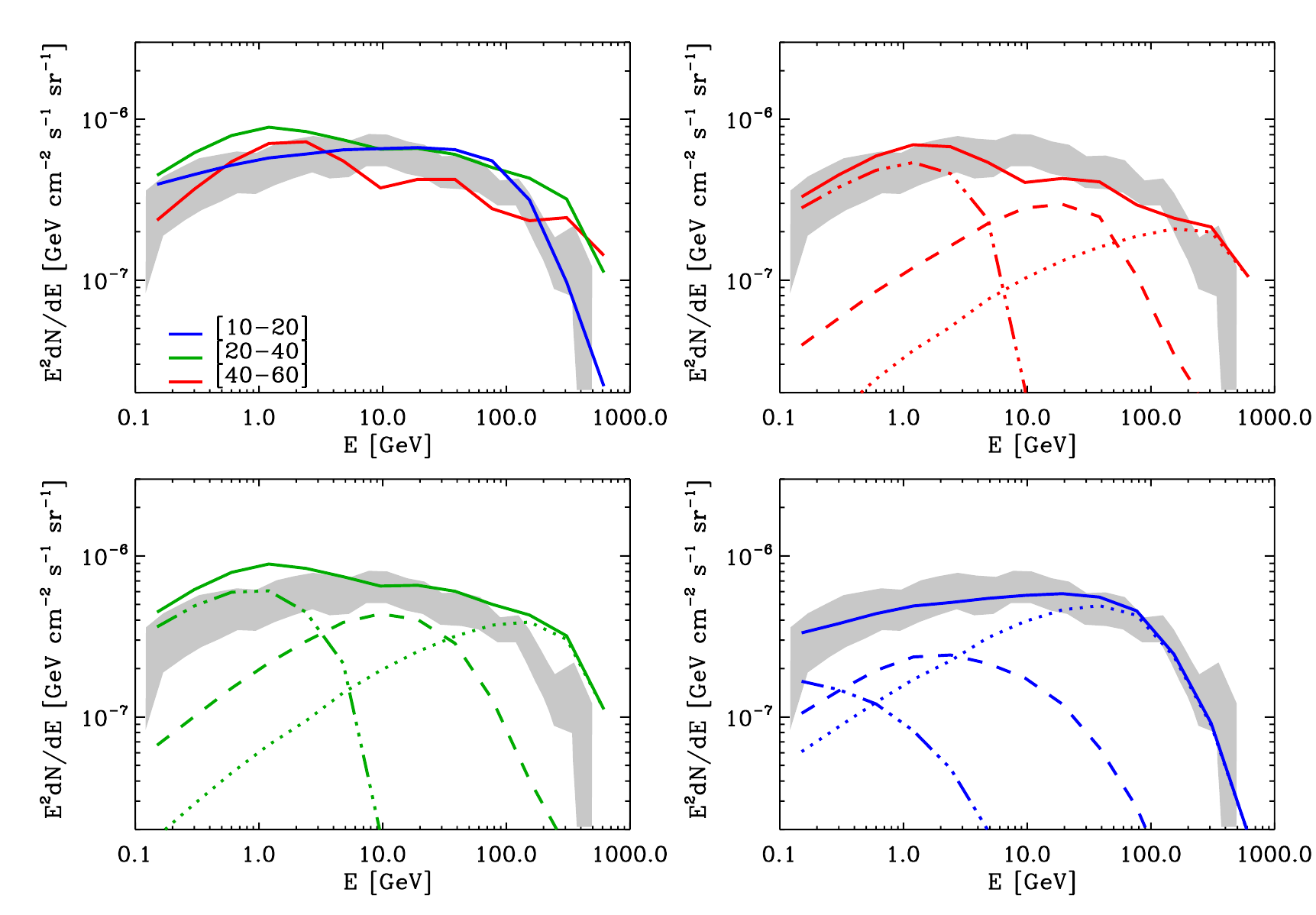} 
\caption{Simulated spectra of the {\it Fermi} bubbles calculated for a longitude range of $l=[-10^\circ, 10^\circ]$ for different latitude bins (top left panel). The other three panels show decomposition of the simulated spectra into different components of the ISRF, namely the CMB (dashed-triple-dotted), IR (dashed), and optical (dotted) radiation field. The grey band represents the observational data of \cite{Ackermann14}. The leptonic jet model successfully reproduced the latitude independence of the observed spectra, including the normalization, overall spectral shape, and the spectral cutoff above $\sim 110$ GeV, despite the complex convolution of CR energies and the latitude-dependent ISRF.} 
\label{fig:spectra}
\end{center}
\end{figure*}

Because of the spatially uniform distribution of $E_{\rm max}$, we expect the gamma-ray spectrum to have a spatially uniform high-energy cutoff at similar energies ($E_{\rm max,obs} \lesssim E_{\rm max}$ for intermediate and high latitudes; see \S~\ref{sec:hints}). In the upper left panel of Figure \ref{fig:spectra} we plot the simulated gamma-ray spectra of the {\it Fermi} bubbles calculated for a longitude range of $l=[-10^\circ, 10^\circ]$ for different latitude bins. For a given longitude and latitude range, the simulated spectrum is computed by projecting the gamma-ray emissivities as a function of energy along finely sampled lines of sight (with resolutions of 0.5 degrees), and then we average the spectra over all the sightlines within the region. Indeed, we find that the simulated spectra for all latitudes exhibit a spectral cutoff at similar energies around several hundreds GeV, consistent with the observed cutoff energy. 

The top left panel of Figure \ref{fig:spectra} also shows that, not only is the high-energy cutoff similar across all latitudes, but the general shape of the simulated spectra are also latitude independent as observed. As discussed in \S~\ref{sec:hints}, the energy of CRe must be slightly higher at higher latitudes because the CMB (optical) photons dominates the ISRF at high (low) latitudes. Here we note that although we did not account for diffusion, CR diffusion in the solar neighborhood is known to be energy dependent,
scaling as $E^{0.3 - 0.6}$ \citep{Strong07}.  Advection expands the bubbles to $\sim$ 6 kpc in $\sim$ 1 Myr;  for CRs to diffuse a comparable distance their diffusivity $D$ would have to be of order $1.1\times 10^{31}$ cm$^2$s$^{-1}$. This is 220 times larger than $D$ for GeV particles ($D\sim 5\times 10^{28}$ cm$^2$s$^{-1}$), but if $D$ scales with energy as $E^{0.5}$, diffusion would dominate advection for $E > 50$ TeV. While diffusion is likely to be only a small effect, it goes in the right direction to explain an excess of high energy particles at large heights. In order to see this effect more clearly, we decompose the spectra for each latitude bin into three components, which are calculated from the CMB, IR, and optical photons in the ISRF. At high latitudes ($b = 40^\circ-60^\circ$), the simulated CRe have an average energy of $\langle E_{\rm e} \rangle\sim 530$\ GeV, and therefore the gamma-ray spectrum has a bump with $E_{\rm bump} \sim 1$\ GeV after up-scattering the CMB photons (see Eq.\ \ref{eq:egamma}). Because for this latitude bin the IC scattering goes from the Thomson regime to the KN regime, one can also see the change in spectral indices from $\sim 1.5$ to $\sim 3$ from low to high gamma-ray energies (see \S~\ref{sec:hints}). For intermediate latitudes ($b = 20^\circ-40^\circ$), the three components make comparable contributions to the spectrum. For the low latitude bin ($b = 10^\circ-20^\circ$), the gamma-ray emission is dominated by IC scattering of the optical starlight, with $E_{\rm bump} \sim 40$\ GeV and an average CR energy of $\langle E_{\rm e} \rangle \sim 40$\ GeV. Because at low latitudes the scattering is in the Thomson limit, the spectral index is $\sim 1.5$ up to the cutoff energy.  

In short summary, we demonstrated that the spectra of the {\it Fermi} bubbles are nearly latitude independent because the CRe from different parts of the bubbles at the present day all originate from the GC where they suffer from fast synchrotron and IC cooling soon after injections. We reproduced the {\it latitude-independent} cutoff energy and spectral shape of the gamma-ray spectra despite the complex convolution of CR energies and the {\it latitude-dependent} ISRF. We also note that the normalizations of the simulated spectra for different latitude bins are comparable to one another (top left panel in Figure \ref{fig:spectra}), indicating the flat surface brightness of the observed bubbles is also recovered. This is quite a remarkable result since one must get the CR distribution right both {\it spatially} and {\it spectrally} in order to successfully reproduce the flat intensity and latitude-independent spectra simultaneously.

\subsection{Constraints on the initial conditions}
\label{sec:constraint}

Because the maximum energy of the CRe at the present day, $E_{\rm max}$, is largely determined by fast cooling of CRe near the GC, it could be used to constrain the initial conditions at injection, including the initial speed of the AGN jets and the energy densities of the ISRF and the magnetic field. In this section we discuss the parameter space allowed to build a successful model, and how it would be influenced by improved measurements of the cutoff energy from future observational data. In deriving these constraints, we assume that no significant re-acceleration of CRs took place near the GC.    

Two criteria need to be satisfied at early stage of the bubble evolution in order to generate a spatially uniform bubble spectrum in the scenario described in \S~\ref{sec:spatial}. First, the initial cooling must be fast enough to act on the jets before they propagate away from the GC. Therefore, the cooling timescale of CRe must be shorter than the dynamical time of the jets, i.e., $\tau_{\rm syn+IC} < \tau_{\rm dyn}$. Using the expression for the synchrotron and IC cooling time (Eq.\ \ref{eq:tsynic}) and the definition of $t_{\rm dyn} \equiv (1\ {\rm kpc})/v_{\rm jet}$, we obtain an upper limit on the initial jet velocity,
\begin{equation}
v_{\rm jet} < 0.065c \left(\frac{u_{\rm tot}}{10^{-11}\ {\rm erg\ cm^{-3}}} \right) \left( \frac{E_{\rm max,0}}{{\rm TeV}} \right),
\label{eq:v_upper}
\end{equation}
where $c$ is the speed of light, $E_{\rm max,0}$ is the characteristic maximum energy of CRe near the GC, $u_{\rm tot} = u_{\rm B}+u_{\rm rad}F_{\rm KN}$ is the summation of the energy density of the magnetic field and the ISRF with the correction factor for the KN effect \citep{Moderski05}. Note that the strengths for both the magnetic field and the ISRF rapidly decay away from the GC, and hence $u_{\rm tot}$ in the above equation represents an average value near the GC (roughly within the central kpc). For the following discussion, we assume $f_{\rm cool} \equiv E_{\rm max}/E_{\rm max,0} = 0.3$ to account for the difference between the characteristic CR energy near the GC ($E_{\rm max,0}$) and that observed today ($E_{\rm max}$).  

Another criterion comes from the fact that the initial cooling cannot be so strong that the energy of the CRe cools below the energy required to produce the observed high-energy cutoff today. In other words, the energy of CRe after the initial cooling losses has to be greater than the maximum energy of the CRe today, i.e., $E > E_{\rm max}$. The CR energy after going through synchrotron and IC losses is given by $E=E_0/(1+\beta t E_0)$ \citep{Kardashev62}, where $E_0$ is the initial CR energy and $\beta = (4/3) (\sigma_T/m_e^2 c^3) u_{\rm tot}$. For very large $E_0$, the CR energy after cooling is approximately 
\begin{equation}
E \sim \frac{1}{\beta t} \sim 2.5\ {\rm TeV} \left( \frac{u_{\rm tot}}{10^{-11}\ {\rm erg\ cm^{-3}}} \right)^{-1} \left( \frac{\tau_{\rm dyn}}{0.018\ {\rm Myr}} \right)^{-1}.
\label{eq:ecool}
\end{equation}
The requirement of $E > E_{\rm max}$ gives a lower limit on the jet speed,
\begin{equation}
v_{\rm jet} > 0.02c \left( \frac{u_{\rm tot}}{10^{-11}\ {\rm erg\ cm^{-3}}} \right) \left( \frac{E_{\rm max}}{300\ {\rm GeV}} \right).
\label{eq:v_lower}
\end{equation}

\begin{figure}[tp]
\begin{center}
\includegraphics[scale=1.0]{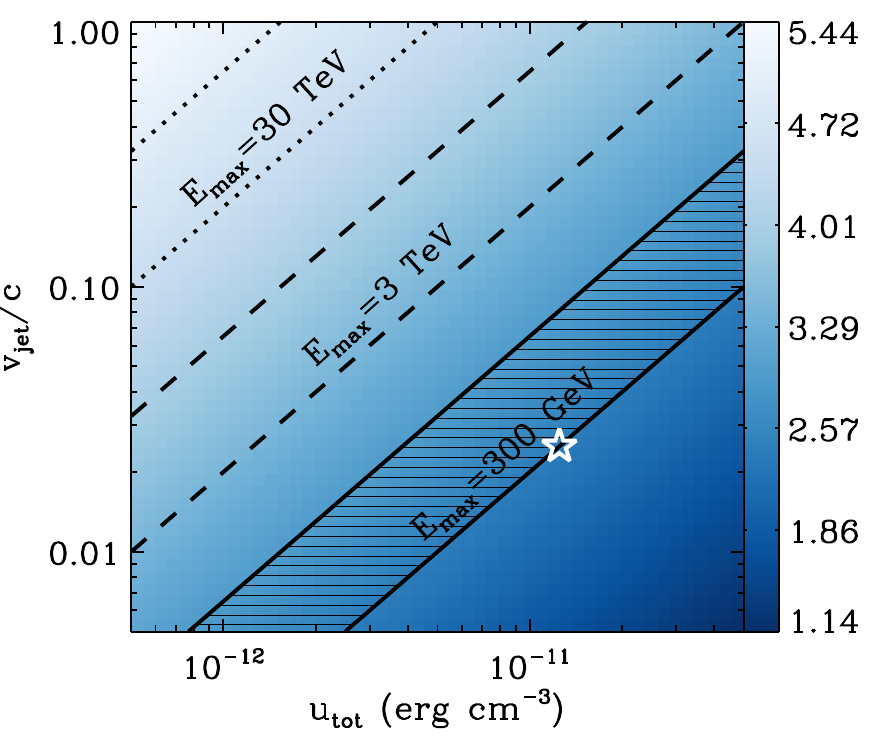} 
\caption{Allowable parameter space for successful models. The $x$ axis represents an average value of the summation of energy densities from the ISRF and magnetic field near the GC, and the $y$ axis is the initial velocity of the jets. The color shows $\log(E_{\rm max}/{\rm GeV})$, where $E_{\rm max}$ is the value of maximum CR energy, for given values of $v_{\rm jet}$ and $u_{\rm tot}$. Parameters within the shaded region satisfy the upper and lower limits of $v_{\rm jet}$ given by Eq.\ \ref{eq:v_upper} and Eq.\ \ref{eq:v_lower}, respectively, assuming $E_{\rm max}=300$\ GeV and $f_{\rm cool}=0.3$ (see the text for definition). The star symbol shows the parameters used in the current simulation. The region bracketed by the dashed and dotted lines is the permitted parameter space assuming $E_{\rm max}=3$\ TeV and 10 TeV, respectively, indicating that future observational limits of $E_{\rm max}$, if bigger than 300\ GeV, would shift the allowable parameter space to the upper-left corner. These constraints are derived assuming there is no significant re-acceleration of CRs near the GC.}
\label{fig:constraints}
\end{center}
\end{figure}

In Figure \ref{fig:constraints} we plot the permitted values of $v_{\rm jet}$ as a function of $u_{\rm tot}$ bracketed by the above two criteria (Eq.\ \ref{eq:v_upper} and \ref{eq:v_lower}) in the shaded region, assuming $E_{\rm max}=300$\ GeV. The color shows the value of $E_{\rm max}$ for given values of $v_{\rm jet}$ and $u_{\rm tot}$ (Eq.\ \ref{eq:ecool}). The parameter set adopted in the current simulation (plotted using the star symbol) lies within the permitted parameter space and is therefore able to successfully reproduce the spatially uniform spectrum of the bubbles. However, this figure illustrates that the solution is not unique. \footnote{Though not unique, the jet parameters adopted in the current simulations have been shown to satisfy many other observational constraints (see Y12 for detailed discussion), in additional to those presented here.} For example, for the current observational constraint of $E_{\rm max} \gtrsim E_{\rm max,obs} \sim 300$\ GeV (near the lower solid line), if we were to use an average energy density for the magnetic field and the ISRF of $u_{\rm tot} = 2\times 10^{-11}\ {\rm erg\ cm^{-3}}$, the initial velocity of the AGN jets would have to be in the range of $0.04c - 0.13c$ in order to have a successful model. Generally speaking, in order to produce CRe with energy $E_{\rm max} \gtrsim 300$\ GeV, the initial jet velocity must be faster (slower) for larger (smaller) initial strength for the magnetic field and ISRF. 

Figure \ref{fig:constraints} also shows that, assuming $u_{\rm tot}$ at the time of injection is not significantly smaller than the value adopted in the current simulation, the required initial velocity of the outflow that transports the CRe must be at least $\sim 0.01c$ or $3000\ {\rm km\ s^{-1}}$. Such a fast speed is easily achievable by AGN jets but not by winds driven by nuclear starburst, for example. Therefore, the mechanism for generating spatially uniform spectrum proposed in this work would not be applicable for models that are based on outflows with lower velocities.    

Finally, we discuss the influence on the allowable parameter space by improved constraints on $E_{\rm max}$ in the future from GeV and TeV observations such as {\it Fermi}, High Altitude Water Cherenkov (HAWC) \citep{HAWC17}, Cherenkov Telescope Array (CTA), Large High Altitude Air Shower Observatory (LHAASO), and the Hundred Square km Cosmic ORigin Explorer (HiScore). In Figure \ref{fig:constraints} we plot the permitted parameter space assuming $E_{\rm max}=3$\ TeV and 30\ TeV (bracketed by the dashed and dotted lines, respectively), assuming a constant $f_{\rm cool}=0.3$. If the constraints from future data finds $E_{\rm max}$ to be greater than 300\ GeV, the average energy density of the magnetic field and the ISRF near the GC, $u_{\rm tot}$, would have to be smaller, and/or the initial jet velocity must be higher, in order to be consistent with the observed $E_{\rm max}$. As the $E_{\rm max}$ gets bigger and bigger, the limits on $v_{\rm jet}$ and $u_{\rm tot}$ become more and more stringent. Therefore, if $E_{\rm max}$ approaches tens of TeV, the leptonic jet model would become less favorable because it is difficult for the jets to avoid cooling and keep the CRe at such high energies, unless significant re-acceleration of CRs occured near the GC to compensate for the cooling losses. If there were CR re-acceleration, it would have similar effects as slower cooling and shift the permitted parameter space shown in Figure \ref{fig:constraints} downward.

\section{Conclusions}
\label{sec:conclusion}


One of the unique features of the {\it Fermi} bubbles is the spatially uniform spectrum, including the spectral shape and the high-energy cutoff above 110\ GeV. Because reproducing the latitude-independent spectrum requires the correct CR distribution both spatially and spectrally, it provides stringent constraints on the theoretical models proposed to explain the origin of the bubbles. In this work, we investigate the spectral evolution of the {\it Fermi} bubbles in the leptonic AGN jet scenario using 3D hydrodynamic simulations that include modeling of the CR spectrum. The simulations are done using the newly implemented CRSPEC module in FLASH, which allows us to track the spectral evolution of CRe due to adiabatic processes and synchrotron plus IC cooling after they are injected with the AGN jets from the GC. Our main findings are summarized as follows. 

(1) The high-energy cutoff in the observed gamma-ray spectrum of the bubbles is a signature of fast synchrotron and IC cooling of CRe near the GC when the jets were first launched. 

(2) After the initial phase of fast cooling near the GC, the dynamical time of the jets becomes the shortest among all other cooling timescales and therefore the CR spectrum is essentially advected with only mild cooling losses. This could explain why the bubble spectrum is nearly spatially uniform, because the CRe from different parts of the bubbles today all share the same origin near the GC at early stage of the bubble expansion. 

(3) The simulated distribution of CR energies, despite being quite uniform, has a slight gradient toward higher energies at higher Galactic latitudes. We show that this is essential for reproducing the latitude-independent shape of the gamma-ray spectrum because the ISRF is dominated by lower-energy CMB photons at high latitudes and optical starlight at low latitudes. 

(4) Because the observed cutoff energy of the gamma-ray spectrum today is closely related to the early phase of fast cooling, it can be used to constrain the initial conditions near the GC, such as the initial speed of AGN jets and the energy density of the magnetic field and the ISRF. The permitted parameter space for building a successful model and its dependence on the future measurements of the cutoff energy are summarized in Figure \ref{fig:constraints}. 

Finally, we note that in addition to the above spectral features, the simulated 3D CR distribution is edge-brightened (Figure \ref{fig:slices}), which is key for recovering the flat surface brightness of the observed bubbles, or the latitude-independent normalization of the observed spectrum (top left panel of Figure \ref{fig:spectra}). It is remarkable that the leptonic jet model predicts the right spatial and spectral distribution of CRe that simultaneously matches the normalization, overall spectral shape, and high-energy cutoff of the observed gamma-ray spectrum and their spatial uniformity. Together with the fact that the microwave haze is more easily explained by the leptonic jet model, we conclude that past AGN jet activity is a likely mechanism for the formation of the {\it Fermi} bubbles. Future data from multi-messenger observations, particularly improved measurements of the cutoff energy of the gamma-ray spectrum by GeV and TeV observatories including {\it Fermi}, HAWC, CTA, LHAASO, and HiScore, will provide crucial verification of the scenario proposed in this work.


\acknowledgments
The authors would like to thank Jay Gallagher for helpful discussion. We also thank Ellen Zweibel and an anonymous referee for useful comments that helped to improve the manuscript. HYKY acknowledges support by NASA through Einstein Postdoctoral Fellowship grant number PF4-150129 awarded by the Chandra X-ray Center, which is operated by the Smithsonian Astrophysical Observatory for NASA under contract NAS8-03060. MR acknowledges support from NASA grant NASA ATP 12-ATP12-0017 and NSF grant AST 1715140. 
The simulations presented in this paper were performed on the Deepthought2 cluster, maintained and supported by the Division of Information Technology at the University of Maryland College Park. FLASH was developed largely by the DOE-supported ASC/Alliances Center for Astrophysical Thermonuclear Flashes at the University of Chicago. Data analysis presented in this paper was partly performed with the publicly available {\it yt} visualization software \citep{Turk11}.


\bibliography{fermi}


\appendix

\section{The CRSPEC module}
\label{appendix}

The evolution of CR particles is described by the diffusion-advection equation \citep{Skilling75},
\begin{eqnarray}
\frac{\partial f}{\partial t} = - {\bm v} \cdot \nabla f + \nabla \cdot (\kappa \nabla f) + \frac{1}{3} (\nabla \cdot {\bm v})p \frac{\partial f}{\partial p}
+ \frac{1}{p^2} \frac{\partial}{\partial p} \left[ p^2 \left( b_l (p) f + D_{pp} \frac{\partial f}{\partial p} \right) \right] + j({\bm x},p),
\label{eq:f}
\end{eqnarray}
where $f({\bm x},p)$ is the isotropic part of the particle distribution function, $\kappa (p)$ and $D_{pp} (p)$ are the diffusion coefficients in spatial coordinates and in momentum space, respectively, $b_l(p)$ describes mechanical and radiative losses (see Eq.\ \ref{eq:bl}), and $j({\bm x},p)$ is the source term accounting for CR injections at shocks or production of secondary particles. In order to save computational costs, the momentum space is divided into relatively sparse $N_p$ logarithmically spaced momentum bins, bounded by $p_0$,...,$p_{N_p}$. The width of the bins on a log scale, $\Delta \log p = \log (p_i/p_{i-1})$, is assumed to be a constant for convenience. The modeled CRs have a finite spectrum with minimum and maximum momenta of $p_{cutL}$ and $p_{cutR}$, respectively, and one has to ensure that $p_0$ and $p_{N_p}$ bracket the evolution of $p_{cutL}$ and $p_{cutR}$. For momentum bins that at least partially intersect with the CR spectrum, the distribution function $f(p)$ is approximated by a piece-wise power-law distribution, 
\begin{equation}
f(p) = f_i \left( \frac{p}{p_L} \right)^{-q_i}, \
p_L = \begin{cases} p_{i-1}, & {\rm if}\ p_{i-1} \geq p_{cutL} \\ 
                                p_{cutL}, & {\rm otherwise} \end{cases} \\ 
\label{eq:fp}
\end{equation}
where $f_i$ and $q_i$ are the normalization and logarithmic slope for the $i$th bin. For momentum bins that do not contain any CRs, i.e., if $p_{i-1} > p_{cutR}$ or $p_i < p_{cutL}$, $f_i$ is assigned to be zero. After integrating over the $i$th momentum bin of Eq.\ \ref{eq:f} multiplied by $4\pi p^2$, we obtain equations for the CR number densities, 
\begin{eqnarray}
\frac{\partial n_i}{\partial t} &=& - \nabla \cdot ({\bm v} n_i) + \nabla \cdot ( \langle \kappa \rangle \nabla n_i ) + \left\{ \left[ \frac{1}{3} (\nabla \cdot {\bm v}) p + \left( b_l(p) + D_{pp} \frac{\partial \log f}{\partial p} \right) \right] 4 \pi p^2 f(p) \right\}^{p_i}_{p_{i-1}} + Q_i, \label{eq:dni_dt} \\
n_i &=& \int^{p_R}_{p_L} 4 \pi p^2 f(p) dp = 4 \pi f_i p^3_L \frac{(p_R/p_L)^{3-q_i}-1}{3-q_i}, \ \label{eq:ni}
p_R = \begin{cases} p_{i}, & {\rm if}\ p_{i} \leq p_{cutR} \\ 
                                p_{cutR}, & {\rm otherwise} \end{cases} \\
\langle \kappa \rangle_i &=& \frac{\int_{p_L}^{p_R} p^2 \kappa \nabla f dp}{\int_{p_L}^{p_R} p^2 \nabla f dp}, \\
Q_i &=& \int_{p_{i-1}}^{p_i} 4\pi p^2 j(p) dp.
\end{eqnarray}
The advection and spatial diffusion terms (i.e., first and second terms in Eq.\ \ref{eq:dni_dt}) are implemented in the same way as in the energy-integrated version of the CR module (Y12). The terms that represent second order Fermi acceleration ($\propto D_{pp}(p)$) and the source term ($Q_i$) are neglected hereafter in order to focus on processes relevant for this paper. The evolution of $n_i$ due to advection in momentum space becomes
\begin{eqnarray}
\frac{\partial n_i}{\partial t} = \left[ b(p) 4 \pi p^2 f(p) \right]^{p_i}_{p_{i-1}}, \label{eq:ni_evol} \\
b(p) \equiv \frac{dp}{dt} = \frac{1}{3} (\nabla \cdot {\bm v}) p + b_l(p), \label{eq:bp}
\end{eqnarray}
where $b(p)$ includes adiabatic compression or expansion as well as other energy loss terms $b_l(p)$ (see Eq.\ \ref{eq:bl}). Note that $b(p)$ is greater (smaller) than zero when CRs are cooling (heating). Integration of Eq.\ \ref{eq:ni_evol} over time gives
\begin{eqnarray}
n_i^{t+\Delta t} &-& n_i^t = - \Delta t (\Phi_i^p - \Phi_{i-1}^p), \label{eq:ni_dt} \\
\Phi_i^p &=& - \frac{1}{\Delta t} \int_t^{t+\Delta t} b(p) 4\pi p^2 f(t',p)|_{p_i} dt',
\end{eqnarray}
where $\Phi_i^p$ is the time-averaged flux evaluated at the cell boundary $p_i$. Using Eq.\ \ref{eq:bp} to rewrite the above equation, we have
\begin{eqnarray}
\Phi_i^p = - \frac{4\pi}{\Delta t} \int_{p_i}^{p_u} p^2 f_j(p) dp, \label{eq:phip} \\
j = \begin{cases} i+1, & {\rm if}\ b(p_i) > 0 \\
                           i, & {\rm if}\ b(p_i) \leq 0, \end{cases} \label{eq:j}
\end{eqnarray}
where $p_u$ is the upstream momentum, which can be solved using the equation 
\begin{equation}
\Delta t = \int_{p_i}^{p_u} \frac{dp}{b(p)}. \label{eq:pu}
\end{equation}
Note that because of the finite CR spectrum, when CRs are cooling ($b(p) > 0$), Eq.\ \ref{eq:phip} is integrated up to $\min(p_u, p_{cutR})$; when CRs are heating ($b(p) < 0$), the upper integration limit is $\max(p_u, p_{cutL})$. 

For CR ions, at each timestep one can solve for $f_i$ and $q_i$ given the updated $n_i$ based on Eq.\ \ref{eq:ni_evol} assuming that the curvature of the spectrum is constant (see \cite{Miniati01} for detailed discussion). However, for fast cooling CRe, this assumption is not valid and therefore it is necessary to employ other constraints. This can be achieved by evolving the CR energy densities, $e_i$, in addition to the number densities, $n_i$. The equations for $e_i$ can be derived by taking one moment of Eq.\ \ref{eq:f}. That is, we multiply Eq.\ \ref{eq:f} by $4\pi p^2 T(p)$, where $T(p) = (\gamma -1) m_e c^2$ is the particle kinetic energy and $\gamma$ is the Lorentz factor, and integrate over the $i$th momentum bin. This yields
\begin{eqnarray}
\frac{\partial e_i}{\partial t} &=& - \nabla \cdot ({\bm v} e_i) + \nabla \cdot ( \langle \kappa_T \rangle_i \nabla e_i) +
\left\{ \left[ \frac{1}{3} (\nabla \cdot {\bm v}) p + b_l(p) \right] 4\pi p^2 f(p) T(p) \right \}_{p_{i-1}}^{p_i} \nonumber \\  
&& - \int_{p_{i-1}}^{p_i} \left[ \frac{1}{3} (\nabla \cdot {\bm v}) p + b_l(p) \right] 4\pi p^2 f(p) \frac{p}{\sqrt{m_e^2 c^2 + p^2}} dp + S_i, \\
e_i &=& \int_{p_L}^{p_R} 4\pi p^2 f(p) T(p) dp = 4 \pi c f_i p_L^4 \frac{(p_R/p_L)^{4-q_i}-1}{4-q_i}, \label{eq:ei} \\
\langle \kappa_T \rangle_i &=& \frac{\int_{p_L}^{p_R} \kappa p^2 (\nabla f) T(p) dp}{\int_{p_L}^{p_R} p^2 (\nabla f) T(p) dp}, \\
S_i &=& \int_{p_{i-1}}^{p_i} 4\pi p^2 j(p) T(p) dp, 
\end{eqnarray}
where Eq.\ \ref{eq:fp} is used and sub-relativistic contribution is ignored in order to derive the last expression in Eq.\ \ref{eq:ei}. Same as before, we neglect the advection, diffusion, and source terms and only focus on the energy transfer in momentum space, i.e., 
\begin{equation}
\frac{\partial e_i}{\partial t} = \left[ b(p) 4\pi p^2 f(p) T(p) \right]_{p_{i-1}}^{p_i} - \int_{p_L}^{p_R} b(p) \frac{4\pi p^3 c f(p)}{\sqrt{m_e^2 c^2 + p^2}} dp. 
\label{eq:ei_evol}
\end{equation}
Using Eq.\ \ref{eq:fp}, the second term in the above equation can be rewritten as $e_i R_i$, where 
\begin{equation}
R_i = \int_{p_L}^{p_R} b(p) \frac{p^{3-q_i}}{\sqrt{m_e^2 c^2 + p^2}} dp \bigg/ \int_{p_L}^{p_R} p^{2-q_i} T(p) dp.
\end{equation}
Integrating Eq.\ \ref{eq:ei_evol} over time gives
\begin{eqnarray}
e_i^{t+\Delta t} \left( 1+ \frac{\Delta t}{2} R_i \right) = e_i^t \left( 1 - \frac{\Delta t}{2} R_i \right) - \Delta t ( \Phi_i^e - \Phi_{i-1}^e ), \label{eq:ei_dt} \\
\Phi_i^e = - \frac{1}{\Delta t} \int_t^{t+\Delta t} b(p) 4 \pi p^2 f_i (t',p) T(p) |_{p_i} d t',
\end{eqnarray}
where $\Phi_i^e$ is the time-averaged flux evaluated at the cell boundary $p_i$. Again, we could rewrite the equation using Eq.\ \ref{eq:bp} and obtain
\begin{equation}
\Phi_i^e = - \frac{4\pi}{\Delta t} \int_{p_i}^{p_u} p^2 f_j(p) T(p) dp,
\end{equation}
where $j$ and $p_u$ are defined in Eq.\ \ref{eq:j} and \ref{eq:pu}, respectively. Similar to Eq.\ \ref{eq:phip}, the upper integration limit is $\min(p_u, p_{cutR})$ and $\max(p_u, p_{cutL})$ when CRs are cooling and heating, respectively. The minimum and maximum momenta of the CR spectrum ($p_{cutL}$ and $p_{cutR}$, respectively) are solved explicitly using 
\begin{equation}
\Delta t = \int_{p_{cut}^{t+\Delta t}}^{p_{cut}^t} \frac{dp}{b(p)}. \label{eq:pcut}
\end{equation}

In the simulations, in addition to hydrodynamic variables, extra $2N_p + 2$ variables are stored for $n_i$, $e_i$, $p_{cutL}$, and $p_{cutR}$. At each simulation timestep, after accounting for CR advection and diffusion, we first convert $n_i$ and $e_i$ into $f_i$ and $q_i$, where $q_i$ could be solved using the following equation for each momentum bin (assuming $p_i \gg m_e c$):
\begin{equation}
\frac{e_i}{n_i p_L c} = \frac{3-q_i}{(p_R/p_L)^{3-q_i}-1} \frac{(p_R/p_L)^{4-q_i}-1}{4-q_i},
\end{equation}
and $f_i$ could be computed directly using Eq.\ \ref{eq:ni} or \ref{eq:ei}. We then update $n_i$ and $e_i$ from $t$ to $t+\Delta t$ using Eq.\ \ref{eq:ni_dt} and \ref{eq:ei_dt}. Finally the minimum and maximum momenta of the CR spectrum are updated using Eq.\ \ref{eq:pcut}. 

In general, in order to achieve numerical accuracy, the simulation timestep $\Delta t$ has to satisfy $| \log \frac{p_u}{p_i}| \leq \epsilon \log \frac{p_i}{p_{i-1}}$, where $\epsilon \leq 1$ is similar to the Courant number. However, when fast cooling CRe are included, $\Delta t \lesssim 0.1\ \tau_{cool}$ should be adopted in order to accurately follow the high-energy end of the spectrum. In the current implementation in FLASH, when timestep constraints due to CR cooling become the most limiting among all relevant timesteps, we subcycle over the CR spectral evolution in order to speed up the computation. Because of the block-structured nature of the AMR architecture in FLASH, neighboring blocks are likely assigned to the same processor. This causes significant load imbalance among processors when a particular region in the simulation domain (e.g., near the GC for the current simulation) suffers from fast CRe cooling. To this end, we pay special attention to load balancing in order to achieve good parallel performance.      


For simulations presented in this paper, only synchrotron and IC losses for CRe are relevant. We refer the readers to \cite{Strong98} for the expressions for other energy losses of CRe and CRp. The energy loss rate of CRe due to synchrotron and IC losses is given by
\begin{equation}
\left( \frac{dE}{dt} \right)_{syn+IC} = \frac{4}{3} c \sigma_T \beta^2 \gamma^2 (u_B + u_{rad} F_{KN}),
\end{equation}
where $\beta = \sqrt{1-v^2/c^2}$, $\gamma=E/(m_e c^2)$ and $u_B$ and $u_{rad}$ are energy densities in magnetic field and the radiation field in units of ${\rm erg\ cm^{-3}}$, respectively. The factor $F_{KN}$ accounts for the reduced IC cross section in the KN regime: 
\begin{equation}
F_{KN} = \frac{1}{u_{rad}} \int_{E_{ph,min}}^{E_{ph,max}} f_{KN} (x) E_{ph} n(E_{ph}) dE_{ph},
\end{equation}
where 
\begin{equation}
f_{KN} (x) \simeq \frac{1}{(1+\Gamma)^{1.5}} \ \ {\rm for} \ \Gamma=\frac{4\gamma E_{ph}}{m_e c^2} \lesssim 10^4
\end{equation}
is an analytical approximation for the general KN formula \citep{Moderski05}.  
The synchrotron and IC cooling time ($\equiv E/(dE/dt)$) is 
\begin{equation}
\tau_{syn+IC} = 0.97\ {\rm Myr}\ \left( \frac{u_B+u_{rad}F_{KN}}{10^{-12}\ {\rm erg\ cm^{-3}}} \right)^{-1} \left( \frac{\gamma}{10^6} \right)^{-1}. \label{eq:tsynic}
\end{equation}
By defining $\hat{p} \equiv p/m_e c$, one can write the energy of CR particles as $E=\sqrt{\hat{p}^2+1} m_e c^2$. Therefore, the momentum loss rate as used in Eq.\ \ref{eq:bp} is related to the energy loss rate by
\begin{equation}
b_l(p) \equiv \frac{dp}{dt} = \frac{\sqrt{\hat{p}+1}}{c\hat{p}} \frac{dE}{dt}.\label{eq:bl}
\end{equation} 

\begin{figure}[tp]
\begin{center}
\includegraphics[scale=0.35]{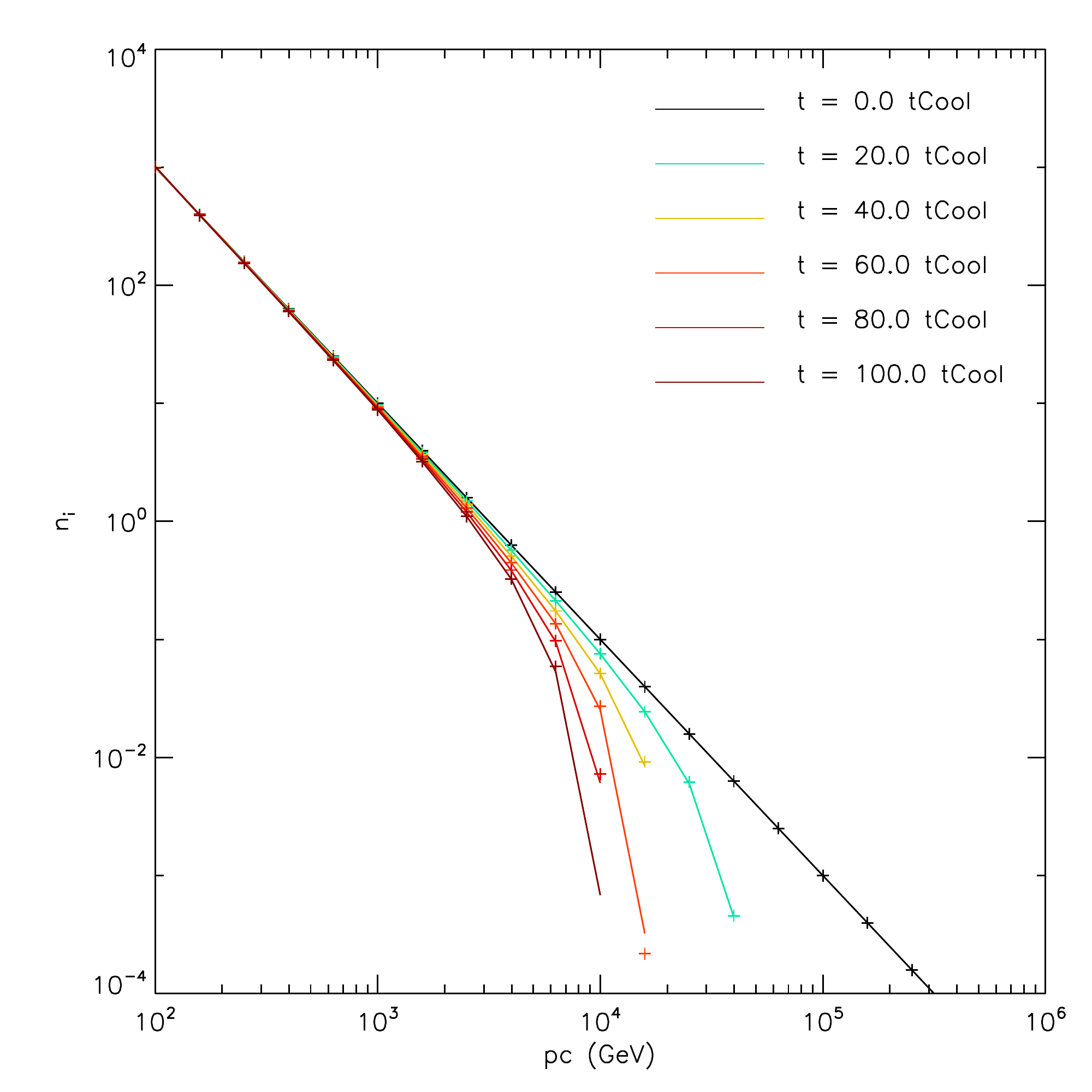} 
\caption{Spectral evolution of CRe due to synchrotron losses. The curves represent the results obtained using the CRSPEC module, and the plus symbols are the analytical solutions. } 
\label{fig:syn}
\end{center}
\end{figure}

Figure \ref{fig:syn} shows a test of the CRSPEC module including synchrotron losses of CRe. The initial CR spectrum ranges from $10^2$ to $10^6$\ GeV with constant spectral indices of $q_i = 5$. The initial spectrum is normalized such that the number density of the first momentum bin $n_1=10^3\ {\rm cm^{-3}}$. The results are in excellent agreement with the analytical solution \citep{Kardashev62}.   

\end{document}